\documentclass[aip,jcp,reprint]{revtex4-1}
\usepackage{graphicx}   
\usepackage{amssymb}
\usepackage{amsmath}
\DeclareMathOperator\erf{Erf}
\DeclareMathOperator\erfi{Erfi}

\begin{document}
\title{Photobleaching of randomly-rotating fluorescently-decorated particles} 
\author{Swadhin Taneja}	
\email{swadhin.taneja@dal.ca}	
\author{Andrew D. Rutenberg}	
\email{andrew.rutenberg@dal.ca}
\affiliation{Dept. of Physics and Atmospheric Science, Dalhousie University, Halifax, Nova Scotia, Canada B3H 4R2}

\date{\today}
\begin{abstract} 
Randomly rotating particles that have been isotropically labeled with rigidly linked fluorophores will undergo non-isotropic (patchy) photobleaching under illumination due to the dipole coupling of fluorophores with light. For a rotational diffusion rate $D$ of the particle and a photobleaching timescale $\tau$ of the fluorophores, the dynamics of this process are characterized by the dimensionless combination $D \tau$. We find significant interparticle fluctuations at intermediate $D \tau$. These fluctuations vanish at both large and small $D \tau$, or at small or large elapsed times $t$.  Associated with these fluctuations between particles, we also observe  transient non-monotonicities of the brightness of individual particles. These non-monotonicities can be as much 20\% of the original brightness. We show that these novel photobleach-fluctuations dominate over variability of single-fluorophore orientation when there are at least $10^3$ fluorophores on individual particles. 
\end{abstract}
\maketitle

\section{Introduction} 
Photobleaching, or the occasional but irreversible loss of fluorescence in individual fluorophores due to illumination, is often an annoyance in biological imaging. Nevertheless, it is used in fluorescence recovery after photobleaching (FRAP) techniques to determine local translational diffusivity. \cite{LippincottSchwartz2001} Cellular copy-number of fluorophores can also be determined by exploiting the fluctuations inherent in the photobleaching process. \cite{Nayak2011, Kim2016} Understanding fundamental physical processes that contribute to observable phenomenology during photobleaching is important for the appropriate application and interpretation of quantitative techniques. 

Fluorophores have a dipolar coupling with the electric field, which means that the fluorophore brightness depends on its orientation with respect to the illumination polarization. \cite{Ha1999} This anisotropy can be exploited to determine the orientation or rotational diffusivity of individual fluorophores. \cite{Ha1998}  With polarized illumination and imaging, polarized fluorescence recovery after photobleaching (PFRAP) can determine slow rotational diffusivity of fluorophores. \cite{Velez1988, Yuan1995} PFRAP relies on the rapid photobleaching of an aligned fraction of fluorophores and the subsequent slow rotation of unbleached fluorophores to provide signal recovery. 

When fluorophores do not rotate, the anisotropic dipolar coupling with the constant illumination beam leads to a non-exponential photobleaching decay of the fluorescent signal with time. \cite{FurederKitzmuller2005}  This is analogous to the non-exponential photobleaching expected in non-uniformly illuminated samples \cite{Berglund2004}, but is due to the non-uniform orientation (random but static) of collections of fluorophores.   

The simultaneous effect of particle rotation and bound fluorophore bleaching has not been previously considered. PFRAP considers the signal recovery due to rotation without further bleaching after rapid photobleaching \cite{Velez1988, Yuan1995}, while non-exponential photobleaching was characterized only for non-rotating particles. \cite{FurederKitzmuller2005}  Understanding the effects of simultaneous particle rotation and fluorophore photobleaching is particularly relevant when multiple fluorophores are bound to an individual particle. 

Fluorescently-labeled polymeric microbeads of various sizes are readily available \cite{Schwartz1998} and can be used to probe the local environment at various length-scales comparable to the particle size. This is particularly interesting within the cellular context, where rotational and translational diffusion can be locally (and distinctly) affected by local membranes \cite{Saffman1975} and crowding \cite{McGuffee2010}.  Typically, a large number of fluorophores are attached to individual particles (e.g. $6 \mu m$ calibration beads have from $10^4$ to $10^6$ fluorophores attached \cite{Vogt1989}).

In this paper, we model ensembles of fluorescently-labeled spherical particles that are randomly rotating under uniform linearly-polarized illumination. In section II, we mathematically solve the temporal evolution of the average angular-distribution of fluorophore orientations and its impact on the apparent particle brightness.  We assume that many fluorophores are rigidly and isotropically bound to and co-rotating with the particles. We find non-exponential photobleaching that extends earlier results for non-rotating particles. \cite{FurederKitzmuller2005}  Some of the calculation details are provided in the appendix, together with their application to PFRAP. \cite{Velez1988, Yuan1995} In section III, we numerically model the stochastic temporal evolution of individual labeled particles for various numbers of fluorophores. We obtain consistent results with the average behavior, but also characterize the interparticle and temporal fluctuations in fluorescence intensity due to random particle rotation.  

\section{Average bleaching with rotation} 
\label{sec:average}

For an ensemble of labeled particles, we first consider the time-dependent distribution function $f(\theta,\phi,t)$ of the orientation of unbleached fluorophores -- where $\theta \in [0,\pi]$ is the polar angle with respect to the polarization axis $\hat{z}$ and $\phi \in [0,2 \pi]$ is the azimuthal angle.  This represents the average behavior of the ensemble, as it evolves in time due to rotational diffusion of the particles together with photobleaching. We consider an initially isotropic ($f=const$) distribution. Photobleaching proceeds through an anisotropic dipole coupling with the linearly  polarized excitation light with the electric field $\vec{E}$ pointing along the $\hat{z}$ axis, while diffusion has an isotropizing effect. 

The dynamical equation for $f$ is 
\begin{equation}  
	\frac{\partial f(\theta,\phi,t)}{\partial t}= D {\nabla}^2 f(\theta,\phi,t) - \frac{cos^2\theta}{\tau} f(\theta,\phi,t), 
	\label{e:dynamics}
\end{equation}  
where $\nabla^2$ represents a spherical Laplacian, i.e. the angular part of the Laplacian that governs rotational diffusion.  $D$ is the rotational diffusion constant that depends on particle size and the local fluid environment, while  $\tau$ is the time-constant controlling photobleaching that  depends on the fluorophore properties and the illumination intensity. (Our timescales, $1/D$ for particle rotation and $\tau$ for fluorophore photobleaching, are both much longer than the de-excitation fluorescence lifetime of single-fluorophore excitation.) The dipolar coupling of the electric field with the fluorophore dipole $\vec{\mu}$ determines the angular factor in the last term since $|\vec{\mu} \cdot \vec{E}|^2 \propto cos^2\theta$.  The dimensionless combination $D \tau$  describes the relative speed of rotational reorientation with respect to photobleaching. 

Because of the dipolar coupling,  azimuthal structure in $f(\theta,\phi)$ does not affect either the average brightness or the bleach rate. Accordingly, we consider only the azimuthal average $\Theta(\theta,t) \equiv \int_0^{2 \pi} d \phi f(\theta,\phi,t)/ 2 \pi$. We can then expand $\Theta$ with respect to a complete set of Legendre polynomials,
\begin{equation}
	\Theta(\theta, t)= \sum\limits_{n=0}^{\infty} a_n(t) P_n(\cos \theta),
	\label{e:LegExp}
\end{equation}
with coefficients $a_n(t)=\frac{2n+1}{2} \int_{-1}^{1} \Theta(x,t) P_n(x) dx$. 

Averaging Eqn.~\ref{e:dynamics} over $\phi$ and substituting Eqn.~\ref{e:LegExp}, we obtain coupled dynamics for the $a_n(t)$, 
\begin{equation}
	\frac{d a_n}{d t}=-D_n a_n - \frac{A_n a_{n-2}+B_n a_{n}+C_n a_{n+2}}{\tau},
\label{e:Numerical}
\end{equation}
where 
\begin{eqnarray}  
	\label{e:ABCD}
	A_n&=& \frac{n(n-1)}{(2n-3)(2n-1)},   \\ \nonumber
	B_n&=& \frac{(2n^2+2n-1)}{(2n-1)(2n+3)},   \\  \nonumber
	C_n&=& \frac{(n+2)(n+1)}{(2n+3)(2n+5)}, \ \ \   \text{and} \\ \nonumber
	D_n&=& Dn(n+1) . 
\end{eqnarray}
(More details of the calculation are provided in Appendix~\ref{a:linear}.)  We see that the diffusive factor $D_n$ is always positive, so that rotational diffusion always decreases $a_n$ with time for $n>0$. The rotational  factors ($A_n$, $B_n$, and $C_n$) mix the Legendre amplitudes $\{ a_n \}$, which can then transiently increase. The corresponding equations for circularly polarized illumination is provided in Appendix~\ref{a:circular}. 

Initially, at $t=0$, we take fluorophores to be isotropically oriented around the particles so that  $a_n(0)=\frac{2n+1}{2}\int_{-1}^{1}\Theta(x,0)P_n(x) dx =\frac{2n+1}{2}\int_{-1}^{1}\frac{P_0(x)}{2} P_n(x) dx=\frac{1}{2}\delta_{n,0}$. We solve Eqn.~\ref{e:Numerical} for the $\{a_n(t)\}$ numerically using a semi-forward Euler method.  

\begin{figure}[t]   
\includegraphics[trim=25mm 30mm 65mm 55mm, clip, width=0.45\textwidth]{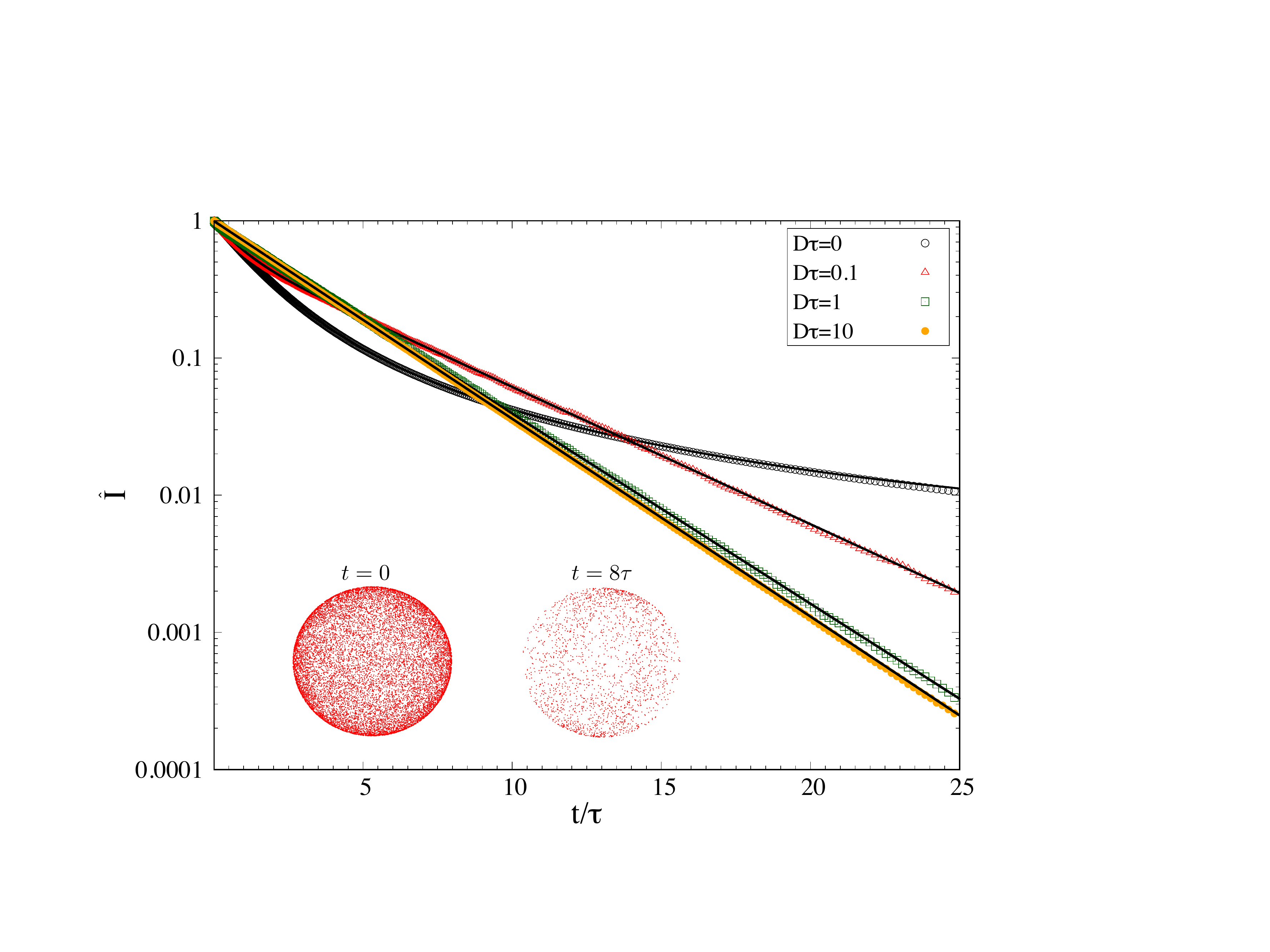}  
\caption{Relative average intensities $\hat{I}$ of unbleached fluorophores on an ensemble of diffusing particles vs scaled time $t/\tau$ for $D \tau = 0, 0.1, 1, 10$ as indicated. The average behavior of the stochastic simulation with $1000$ particles is shown with points, while the numerical solution of Eqn.~\ref{e:BleachIexact} is shown with corresponding lines. Exponential decay is observed for larger $D \tau$, and correspond to straight lines on this semi-log scale. Inset: example of unbleached fluorophores radially oriented around a single spherical particle with $D \tau=0.25$ with $t=0$ or $8 \tau$ as indicated. Initially there are $N=15000$ fluorophores. } 
\label{f:Intensity}
\end{figure}

Under ongoing linearly-polarized illumination, the remaining fluorophores will fluoresce. The time-dependent average intensity per fluorophore, over the ensemble of particles, is given by 
\begin{equation}
	I(t) = \int_{-1}^{1} x^2 ~\Theta(x,t) dx = \frac{4}{15} a_2(t)+ \frac{2}{3} a_0(t).
\label{e:BleachIexact}
\end{equation}
We note that $I(0)=1/3$, where unity corresponds to all fluorophores aligned with $\theta=0$. For this paper, we will show the relative intensity to $t=0$, i.e.
\begin{equation}
	\hat{I} \equiv I(t)/I(0).
\end{equation}
The lines in Fig.~\ref{f:Intensity} show the average relative intensity $\hat{I}$ vs the scaled time $t/\tau$. Exponential bleaching is recovered for larger $D \tau$ values, where the rapid rotational diffusion isotropizes the system.  Non-exponential photobleaching is seen for smaller $D \tau$, consistent with earlier reports at $D \tau = 0$. \cite{FurederKitzmuller2005}

\section{Stochastic rotation and bleaching} 
\label{sec:stochastic}

Underlying the average behavior of an ensemble of particles (described in Section~\ref{sec:average}) is the stochastic behavior of individual particles that randomly rotate and individual fluorophores on those particles that randomly photobleach. Here, we consider this behavior through the stochastic simulation of individual particles and fluorophores. This allows us to consider both the variability between particles at a given time, but also the variation of the brightness of an individual particle with time. By including individual fluorophores, we can also assess when fluctuations due to random rotation and bleaching would be masked due to random initial placement of a small number of initial fluorophores $N$.

We investigate the behavior of isotropically labeled particles that are randomly rotating (with orientational diffusion constant $D$) and that each have $N$ attached fluorophores that rigidly co-rotate with the particle and randomly photo-bleach with rate $\Gamma_i = \cos^2{\theta_i}/\tau$ for the $i$th fluorophore that has polar angle $\theta_i$. Using a timestep of $\Delta t =0.001$ we have implemented small random rotations in each $\Delta t$, consistent with $D$, to all of the fluorophores attached to a given particle. We have allowed individual fluorophores to bleach with probability $p_i= \Gamma_i \Delta t \ll 1$. 

The result is illustrated in the inset of Fig.~\ref{f:Intensity} for $D \tau = 0.25$ at two times as indicated. We have used a spherical particle with radially-oriented fluorophores for illustrative purposes, but equivalently we have shown the fluorophore orientations independently of the particle shape. The initial distribution is isotropic, or uniform on the sphere, at $t=0$.  Some amount of fluctuation is apparent at $t=8 \tau$, arising from the ongoing bleaching in combination with the random rotation of the particle. 

The intensity is given by $I(t) = \sum cos^2\theta_i/N$, where the sum is over the unbleached fluorophores. The initial average intensity is $I(0)=1/3$, as before.  We plot the average relative intensity $\hat{I}$ vs $t/\tau$ as points in Fig.~\ref{f:Intensity}. The average of the single particle stochastic simulations agrees well with the lines showing the calculations of the ensemble average from Section~\ref{sec:average}, as expected.

\begin{figure}[ht]   
\includegraphics[clip,trim=15mm 15mm 15mm 15mm, width=0.5\textwidth]{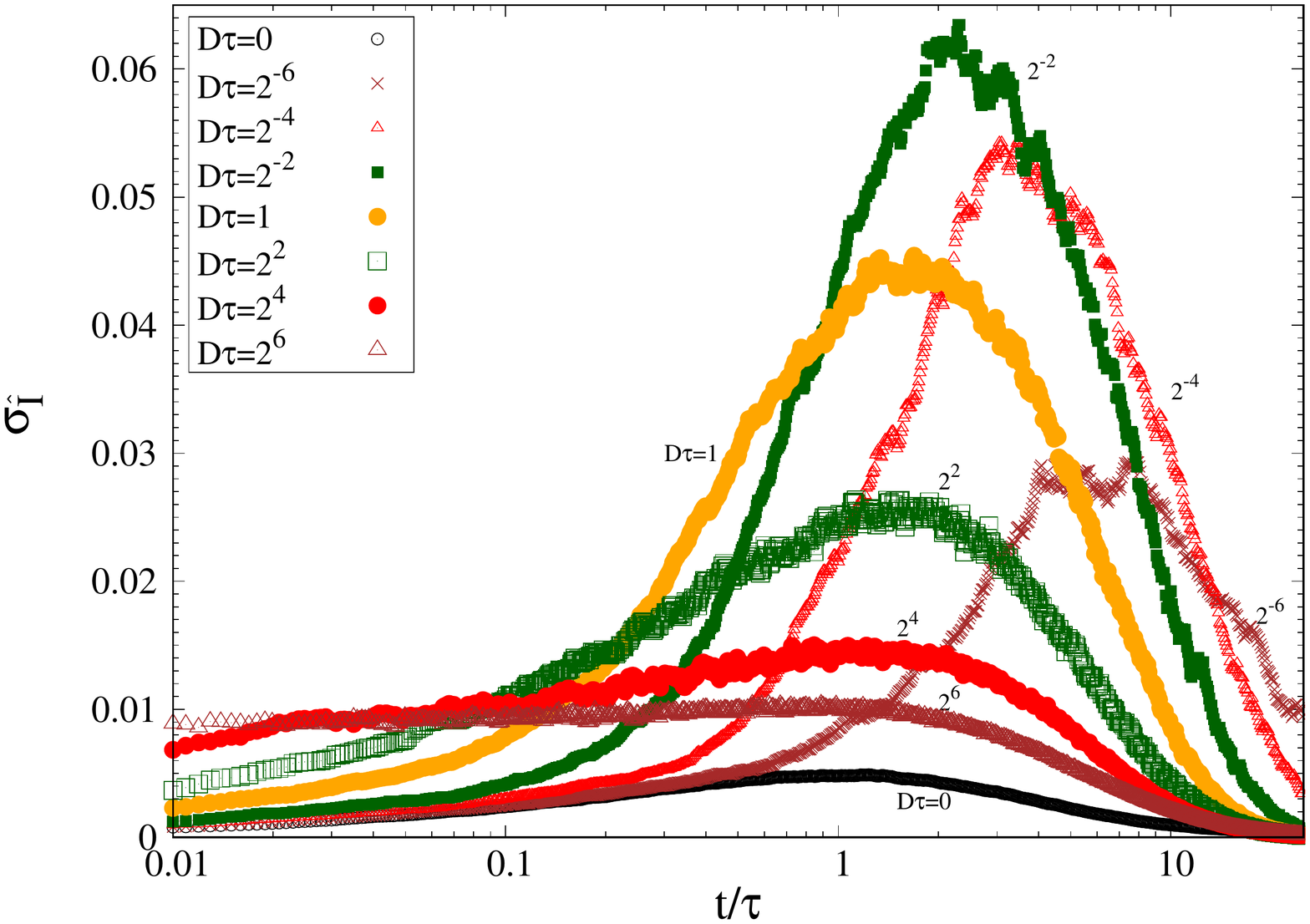}
\caption{Standard deviation between particles of relative intensities $\sigma_{\hat{I}}$ vs scaled time $t/\tau$, for various values of $D \tau$ as indicated by the legend. At $t=0$, $N=10000$ fluorophores were randomly oriented on each particle, which then was allowed to randomly rotate with diffusivity $D$. The fluorophores rigidly rotate with each particle. Stochastic photobleaching of each fluorophore occurs at rate $\Gamma_i = \cos^2 \theta_i/ \tau$, where $\theta_i$ is the polar angle of that fluorophore with respect to the illumination polarization. $1000$ particles were simulated. At $t/\tau \approx 1$ there is a peak in the interparticle fluctuations, which is largest for intermediate values of $D \tau \approx 0.25$.} 
\label{f:StdDev}
\end{figure}

The variability between individual particles is captured by the standard deviation $\sigma_{\hat{I}} = \sqrt{\langle \hat{I}^2 \rangle - \langle \hat{I} \rangle^2}$ of the relative bleach intensity. In Fig.~\ref{f:StdDev} we show $\sigma_{\hat{I}}$
 vs $t/\tau$ for $D \tau =0, 2^{-6}, 2^{-4}, 2^{-2}, 1 , 2^2, 2^4$, and $2^6$, as indicated. With a large number of fluorophores, as $N \rightarrow \infty$, we do not expect any fluctuations in the limits of early times $t/\tau \rightarrow 0$ or for non-rotating particles when $D \tau=0$. The small non-zero fluctuations in these limits result from the finite $N=10000$ number of initial fluorophores. These contributions are small compared to the fluctuations seen at the peak at approximately $t/\tau \approx 1$. The peak is highest for intermediate values of $D \tau \approx 0.25$. These peak fluctuations arise from the different random rotations of individual particles.

\begin{figure}[ht]   
\includegraphics[clip, angle=270,width=0.45\textwidth]{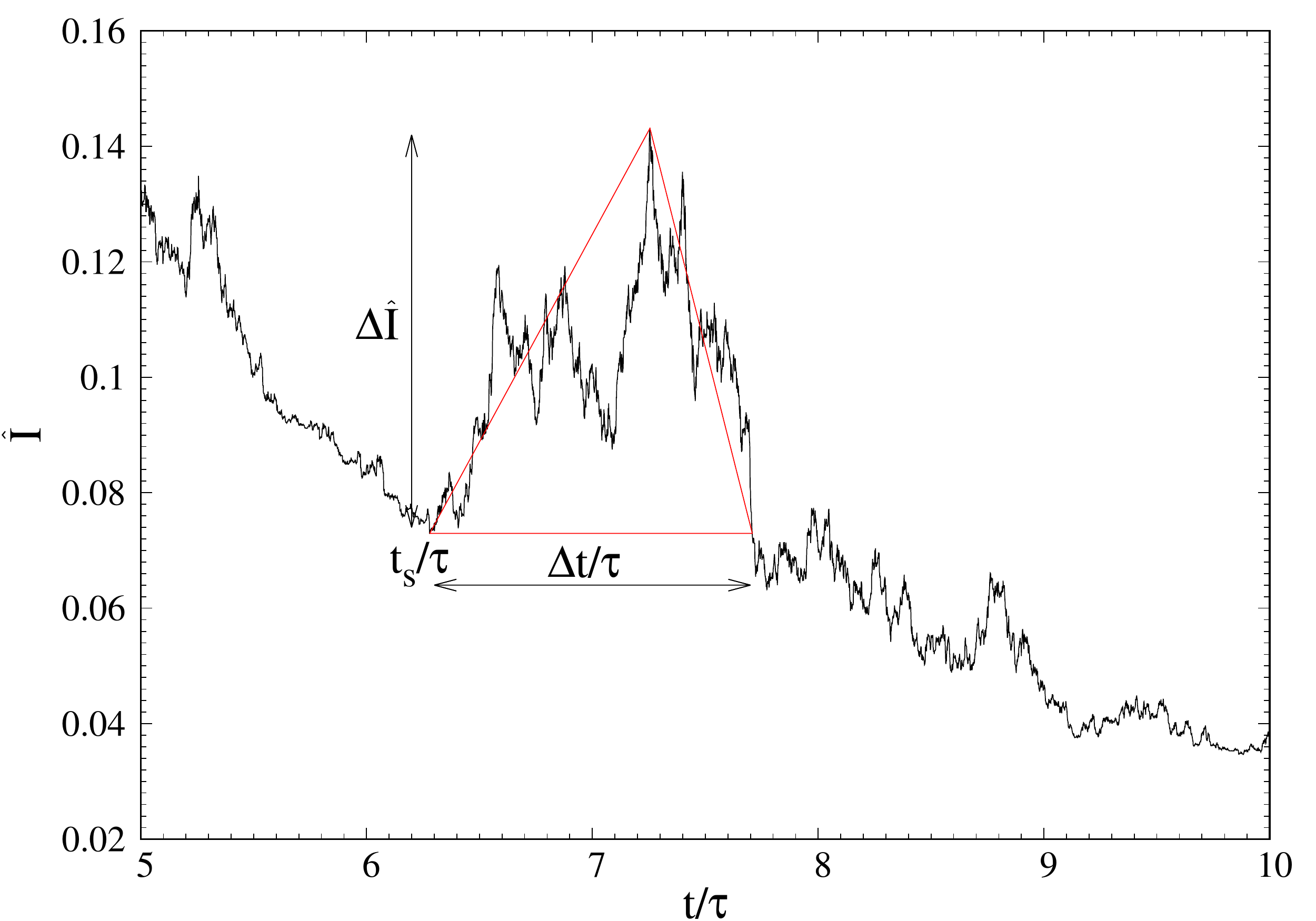}
\caption{For one particle, a portion of the trace of the relative intensity $\hat{I}$ vs. scaled time $t/\tau$ is shown. Here, $D \tau = 1/64$ and $N=1000$. Significant transient increases (non-monotonicities) in $\hat{I}$ are clearly observable, and can be characterized by their magnitude $\Delta \hat{I}$, duration $\Delta t$, and start time $t_s$ -- as indicated.  These non-monotonicities are observed for all $D \tau > 0$.} 
\label{f:Trace}
\end{figure}
 
To better understand the origin of the fluctuations between particles, we considered the relative intensity $\hat{I}$ vs. $t/\tau$ for individual particles. Part of a single-particle trace is shown in Fig.~\ref{f:Trace}. It is apparent that the signal is both stochastic and non-monotonic. These increases of $\hat{I}$ for single particles are due to the rotation of unbleached orientations into alignment with the illumination field, which is the single-particle and continuous-illumination analogue of the PFRAP process (see Appendix~\ref{a:PFRAP} for the average behavior of PFRAP).  We characterize non-monotonic segments, as illustrated by the red triangle in Fig.~\ref{f:Trace}, by a start time $t_s$, an increase of relative intensity $\Delta \hat{I}$, and a duration $\Delta t$. 

\begin{figure}[t]   
\includegraphics[clip, angle=270, width=0.45\textwidth]{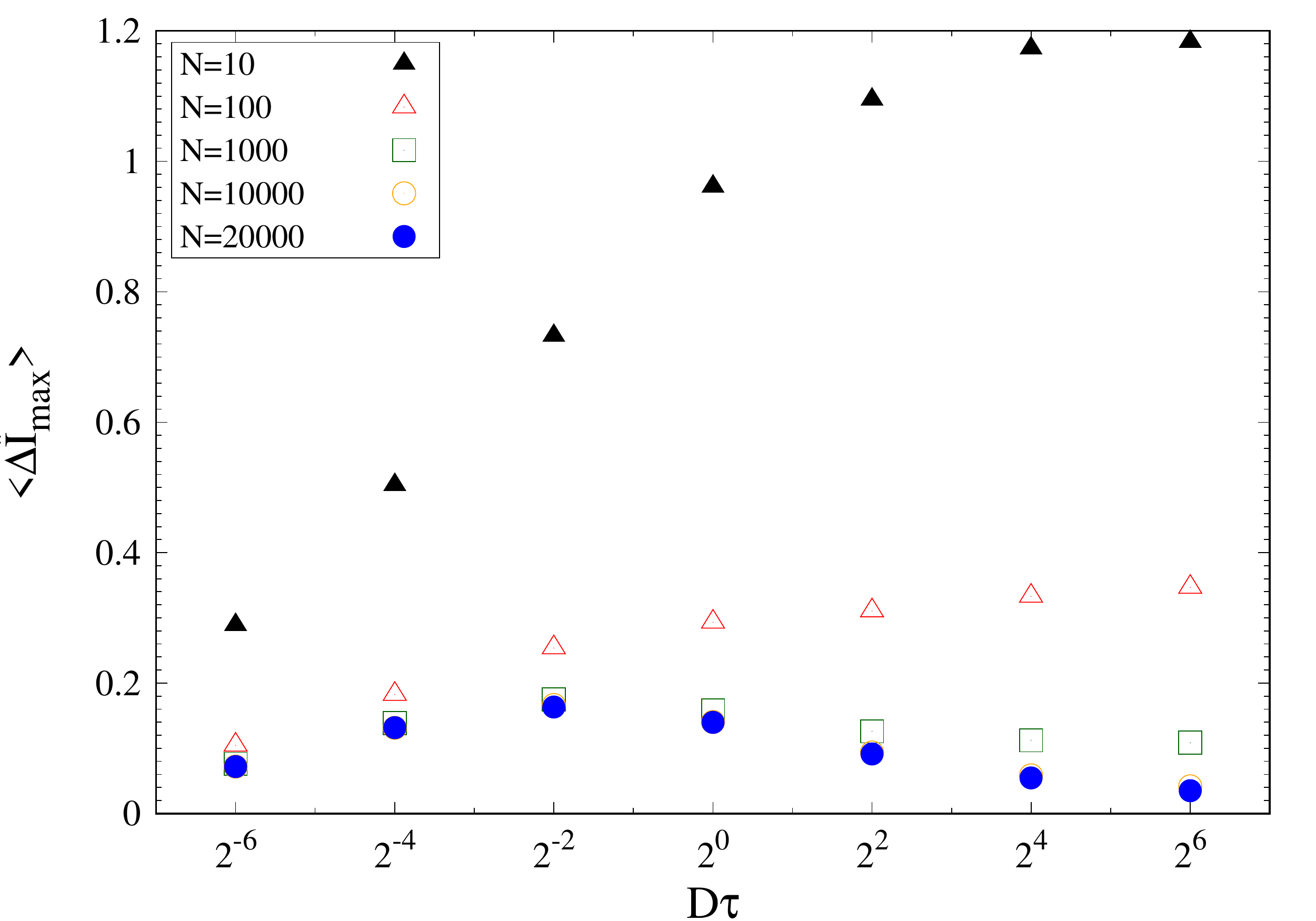}
\caption{The maximal non-monotonicity for each particle $\Delta \hat{I}_{max}$ was averaged and is shown plotted vs the dimensionless $D \tau$ for various numbers of initial fluorophores $N$, as indicated in the legend. For $N \gtrsim 1000$ the non-monotonicities are approximately independent of $N$, and maximal at approximately $D \tau = 0.25$. The magnitudes of maximal non-monotonicities are of the same order as the average interparticle fluctuations in Fig.~\ref{f:StdDev}.} 
\label{f:deltaI}
\end{figure}

We found the non-monotonicity of individual particle intensities surprising, and have characterized its dependence on $D \tau$. We have recorded the maximum absolute increase $\Delta I_{max}$, the corresponding start $t_s$, and duration $\Delta t_{max}$, for $1000$ particle traces. In Fig.~\ref{f:deltaI} we plot the average $<$$\Delta I_{max}$$>$ vs $D \tau$ (note the log-scale). Statistical error bars of the means are smaller than the point sizes. As indicated by the legend, we show the results for various number of fluorophores per particle $N$. For $N=10$ the average intensity increase is larger than $I(0)$ for larger $D \tau$ -- reflecting the spontaneous anisotropy of initial fluorophore orientations with smaller $N$. For larger $N$ the intensity increases are smaller, and for $N \gtrsim 1000$, a maximum of $\langle \Delta \hat{I}_{max} \rangle \approx 0.2$ is apparent at intermediate values of $D \tau \approx 0.25$. This maximum is not caused by random fluorophore placement but by the random rotation (and anisotropic photobleaching) of individual particles, as indicated by the lack of significant $N$ dependence for $N \gtrsim 1000$. 

\begin{figure}[t]   
\includegraphics[clip, angle=270, width=0.45\textwidth]{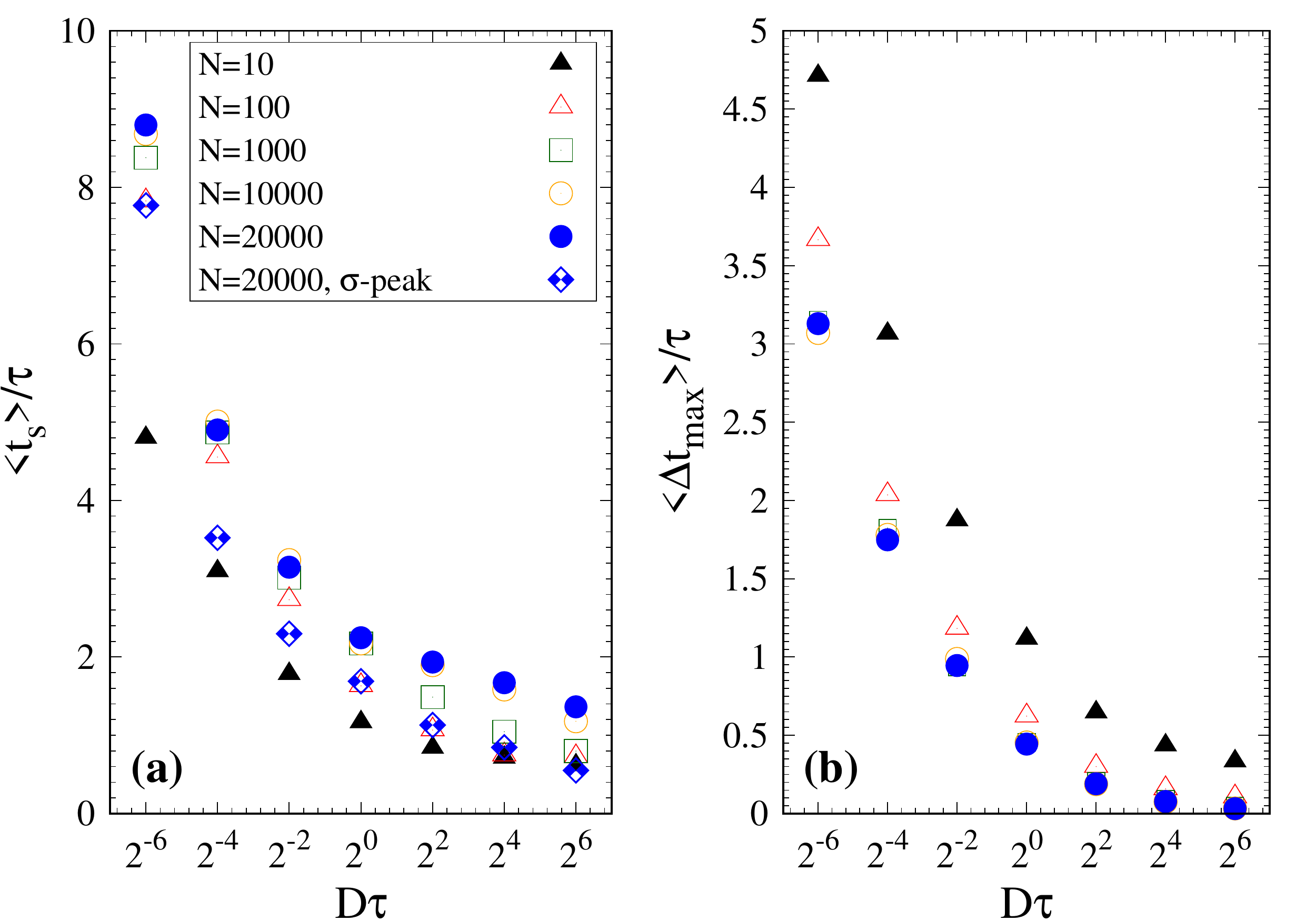}
\caption{The average scaled start time of the maximal non-monotonicity $\langle t_s \rangle/\tau$ vs $D \tau$ is shown in (a), while the average scaled duration $\langle \Delta t_{max} \rangle/\tau$ vs $D \tau$ is shown in (b), for various values of $N$ as indicated by the legend in (a).  Also shown is the timing of the maximal interparticle fluctuation from Fig.~\ref{f:StdDev}. We see that the largest non-monotonicities occur earlier and do not last as long at larger $D \tau$.} 
\label{f:timing}
\end{figure}

The timing and duration of the non-monotonicities are explored in Fig.~\ref{f:timing} and its inset, respectively. We see that at larger $D \tau$ values the largest non-monotonicities occur earlier and do not last as long.   We have also plotted the approximate peak timing of $\sigma_{\hat{I}}$ with blue-diamonds for $N=20000$. We see that the timing of maximal intensity fluctuations between particles is quite close to the timing of maximal non-monotonicities. This implies that non-monotonicities are a significant contribution to the fluctuations between particles. 

\section{Discussion} 

In this work, we have considered isotropically labeled fluorescent particles that rotate with diffusivity $D$ while attached fluorophores are photobleached with timescale $\tau$. The bleach dynamics are controlled by the dimensionless combination $D \tau$.  

We have found that the average bleach dynamics are non-exponential for $0 \leq D \tau < \infty$. We have further characterized the fluctuations between particles, and found that they are maximal at intermediate values of $D \tau \approx 0.25$. By considering individual particles, we found significant random non-monotonicities of their brightness --- approximately corresponding to the maximal fluctuations observed. 

These non-monotonicities and fluctuations are due to the random rotation of unbleached fluorophores into alignment with the polarization of the excitation illumination. This effect is analogous to  polarization recovery after photobleaching (PFRAP) \cite{Velez1988, Yuan1995}, though that involves rapid photobleaching followed by diffusional recovery while this involves simultaneous and continuous photobleaching and diffusion. 

A finite number $N$ of fluorophores will lead to temporal photobleach fluctuations as an $O(\sqrt{N})$ effect \cite{Nayak2011}, and another $O(\sqrt{N})$ effect is expected due to stochasticities in an initial uniform random fluorophore orientation. The effect described in this paper is an $O(N)$ effect that is not due to the finite number of fluorophores. Rather it is due to the random rotation of individual particles, and the rotation of its attached fluorophores along with it. We have found that the $O(N)$ rotational effect dominates over the $O(\sqrt{N})$ effects for $N \gtrsim 1000$.  

We have found that the interparticle fluctuations are largest when $t \approx \tau$ and when $D \tau \approx 0.25$. The later can be easily adjusted since $\tau$ is inversely proportional to the illumination intensity. We only considered linearly polarized illumination, since we expect greater fluctuations than for the more isotropic circularly polarized (or unpolarized) light.  Analogously, PFRAP is only observable for linearly polarized light -- not circularly (see Appendix~\ref{a:PFRAP}).

More broadly, we have identified a new mechanism that contributes to the phenomena of non-exponential photobleaching and of fluorescence fluctuations. There are other sources of non-exponential photobleaching of collections of particles, including depth-extinction \cite{Rigaut1991} and non-uniform illumination \cite{Berglund2004}. Non-exponential bleaching can also be observed for individual fluorophores with multiple internal states  \cite{Berezhkovskii1999}.  Our mechanism of coupled particle rotation and photobleaching generalizes earlier $D=0$ results \cite{FurederKitzmuller2005}. The tunability of our effect with $D$ distinguishes it from other mechanisms of non-exponential photobleaching \cite{Rigaut1991, Berglund2004, Berezhkovskii1999}.

There are also many other sources of fluctuations for fluorophore-associated particles, including blinking and the random orientation and bleaching of a finite number $N$ of fluorophores.  These are typically $O(\sqrt{N})$ effects. Our study adds an $O(N)$ effect due to simultaneous rotation and photobleaching, and this allows it to be distinguished from e.g. blinking or other single-fluorophore effects. Nevertheless, the maximal scale of fluctuations that we have identified is on the order of 5-20\% (see Figs.~\ref{f:StdDev} and \ref{f:deltaI}). Accordingly, we do not anticipate that significant corrections will be needed in correlation spectroscopy techniques \cite{Kolin2007}, which also are more focused on oligomers with $N \lesssim 10$. 

Our model particles are isotropically labeled with many rigidly-bound fluorophores.   How realistic are these ideals with respect to the non-exponential photobleaching and fluctuation phenomena we have characterized? 

Non-spherical nanoparticles, or particles with oriented crystalline or chemically patchy surfaces, would have significant anisotropy in the initial fluorophore orientation. For the average photobleach dynamics (Sec.~\ref{sec:average}) this would introduce non-zero $\{ a_n \}$ for $n>0$, and so modify $\hat{I}(t)$ but not the qualitative observation of non-exponential photobleaching that depends on $D\tau$.  For the fluctuations between particles and in the time evolution of the brightness of single particles (Sec.~\ref{sec:stochastic}) initial anisotropies would likely dominate the fluctuations, much as initial anisotropies introduced by small ($N \lesssim 100$) numbers of bound fluorophores do (Fig.~\ref{f:deltaI}).

Spherical polymeric microbeads are good candidate particles for having isotropically bound fluorophores. We can estimate the minimum particle size by requiring a typical $10nm$ fluorophore separation and $N \gtrsim 1000$ fluorophores per particle. For surface labeled microbeads, we would require a diameter $ d \gtrsim 1 \mu m$. For volume labeled microbeads, a diameter of $d \gtrsim 200 nm$ should suffice. For volume-labeled beads, fluorophore orientation should be independent of bead shape -- so perfectly spherical beads may not be required for isotropy.

Flexible linkers between fluorophores and particles would decrease both anisotropic fluctuations and non-exponential photobleaching. This has been characterized for PFRAP. \cite{Velez1988} Nevertheless, various approaches can minimize such wobble. Multiple single bonds or double bonds between the fluorophore and particle will minimize their relative rotational freedom (see \cite{Rocheleau2003, Beausang2013}). Fluorophores within a glassy matrix, such as within a polymeric microbead, also exhibit limited wobble. \cite{Bhattacharya2016}

With volumetric fluorescent labeling within polymeric microbeads, the ideal conditions of our model should be accessible. Such microbeads of various sizes can separately probe local rotational and translational diffusion at length-scales comparable to the particle size. More generally, we have identified a novel mechanism, of random particle rotation and fluorophore bleaching, for both non-exponential photobleaching and for interparticle and temporal fluctuations in brightness.  This mechanism will contribute to these phenomena even in non-ideal conditions.  We expect qualitatively similar results (though a smaller effect) for circularly polarized illumination.

\acknowledgements 
We thank the ACENET and Westgrid for computational resources, within the Compute Canada federation. ADR thanks the Natural Sciences and Engineering Research Council (NSERC) for operating grant RGPIN-2014-06245, and John Bechhoefer for discussions. 

\appendix

\section{Average Bleach Calculation Details}
\subsection{Linear polarization}
\label{a:linear}
For linear polarization, we start with Eqn.~\ref{e:dynamics}. Applying this to $\Theta(\theta,t)$ and averaging over $\phi$, we obtain:
\begin{equation}
	\frac{\partial \Theta(\theta, t)}{\partial t}= D \frac{1}{\sin\theta}\frac{\partial}{\partial\theta}\left(\sin\theta\frac{\partial \Theta((\theta, t)}{\partial\theta}\right) - \frac{\cos^2\theta}{\tau} \Theta(\theta, t).
	\label{e:thetalin}
\end{equation}
Expanding $\Theta$ in Legendre polynomials $P_n$ and using $\frac{\partial}{\partial x}\left((1-x^2)\frac{\partial P_l(x)}{\partial x}\right)=-l(l+1) P_{l}(x)$, we obtain
\begin{eqnarray}
	\frac{\partial a_n(t)}{\partial t}  = &-& Dn(n+1) a_n(t) \\ \nonumber
		  &-& \frac{(2n+1)}{2} \frac{1}{\tau}\sum\limits_{l=0}^{\infty} a_l(t) \int_{-1}^{1}  x^2 P_n(x) P_l(x) dx.
\end{eqnarray}
Using the identity 
\begin{align*}
\int_{-1}^{1} x^2  P_n(x) P_l(x)dx = 
\begin{cases} \frac{2(l+1)(l+2)}{(2l+1)(2l+3)(2l+5)} & \text{for $n=l+2$}  \\ \\ 
 \frac{2(2l^2+2l-1)}{(2l-1)(2l+1)(2l+3)} & \text{for $n=l$} \\ \\
 \frac{2l(l-1)}{(2l-3)(2l-1)(2l+1)} & \text{for $n=l-2$} 
\end{cases}
\end{align*}
then leads to Eqn.~\ref{e:Numerical}.

\subsection{Circular polarization}
\label{a:circular}
Here, we show the calculation for the average intensity is modified for circularly polarized light. Because of the less anisotropic illumination, we expect smaller fluctuations for the circular polarization case. The dynamics are  
\begin{equation}
	\frac{\partial \Theta(\theta, t)}{\partial t}= D \frac{1}{\sin\theta}\frac{\partial}{\partial\theta}\left(\sin\theta\frac{\partial \Theta((\theta, t)}{\partial\theta}\right) - \frac{\sin^2\theta}{2\tau} \Theta(\theta, t).
	\label{e:thetacirc}
\end{equation}
and we obtain 
\begin{eqnarray}
\frac{\partial a_n(t)}{\partial t}=&-&Dn(n+1) a_n(t)- \frac{1}{2\tau} a_n(t)   \\
	&+& \frac{1}{2\tau}  \left[A_n a_{n-2}(t)+B_n a_{n}(t)+C_n a_{n+2}(t)\right], \nonumber
\end{eqnarray}
also using Eqn.~\ref{e:ABCD}.  $I(t)$ is then given by Eqn.~\ref{e:BleachIexact}.

\subsection{Polarized Fluorescence Recovery After Photobleaching}
\label{a:PFRAP}
PFRAP involves rapid photobleaching followed by rotational recovery. \cite{Velez1988, Yuan1995}  In our approach, rapid photobleaching corresponds to $D=0$.  Subsequent rotational recovery corresponds to $\tau \approx \infty$. 

\subsubsection{PFRAP Linear polarization}
For linear polarization, we can solve Eqn.~\ref{e:thetalin} (with $D=0$) directly by changing variables to $x \equiv \cos(\theta)$. Then $\dot\Theta(x,t) = -x^2/\tau \Theta(x,t)$ and the solution after rapid bleaching for an interval $\Delta t$ is $\Theta(x,\Delta t)=\Theta(x,0) \exp(-x^2 \Delta t/\tau)$ where $\Theta(x,0)=1/2$. 

From Eqn.~\ref{e:BleachIexact} only the $a_0$ and $a_2$ components are relevant to the subsequent rotational recovery (starting at $t=0$, after the rapid bleach). From the Legendre function orthonormality, we have $a_n(0)=\frac{2n+1}{2}\int_{-1}^{1}\Theta(x,0)P_n(x) dx$. This gives
\begin{eqnarray}
	a_0(0) &=& \frac{ \sqrt{\pi} \erf{\sqrt{\alpha}}}{ 4 \sqrt{\alpha}} \\
	a_2(0) &=& \frac{5}{16 \alpha^{3/2}} \left[ -6 \sqrt{\alpha} \exp(-\alpha) + (3-2 \alpha) \sqrt{\pi} \erf{\sqrt{\alpha}} \, \right] \nonumber
	\label{e:linsolution}
\end{eqnarray}
where $\alpha \equiv \Delta t/\tau$ and $\erf(x)$ is the error function.

From Eqn.~\ref{e:thetalin} with $\tau=\infty$ (no significant bleaching during rotational recovery), $a_0$ is time independent, while $a_2(t) = a_2(0) \exp(-6 D t)$. Interestingly $a_2(0) \leq 0$, so the rotational recovery is entirely due to the decay of $a_2(t)$. We find the numerical maximum $a_{2,max}(0) \simeq -0.353$ (with $a_{0,max} \simeq 0.221$) at $\alpha \simeq 3.97$. From Eqn.~\ref{e:BleachIexact} this gives a relative rotational recovery of $64\%$ with PFRAP independent of $D \tau$, which is significantly larger than the $\approx 20\%$ non-monotonicity exhibited in Fig.~\ref{f:deltaI} at $D \tau \approx 0.25$ with continuous bleaching and rotation. 

\subsubsection{PFRAP Circular polarization}
Here  $\dot\Theta(x,t) = -(1-x^2)/(2 \tau) \Theta(x,t)$ and the solution after rapid bleaching for an interval $\Delta t$ is $\Theta(x,\Delta t)=1/2 \exp(-(1-x^2) \Delta t/(2\tau))$. We then obtain
\begin{eqnarray}
	a_0(0) &=& F(\sqrt{\alpha/2})/\sqrt{2 \alpha} \\ 
	a_2(0) &=& \frac{5}{4 \alpha^{3/2}} \left[ 3 \sqrt{\alpha} -\sqrt{2} (3+\alpha) F(\sqrt{\alpha/2}) \, \right] \nonumber
	\label{e:circsolution}
\end{eqnarray}
where $\alpha \equiv \Delta t/\tau$ and the Dawson function $F(x) \equiv  e^{-x^2} \sqrt{\pi} \erfi(x)/2$, where $\erfi$ is the imaginary error function.

Interestingly, $a_2(0) > 0$ for all $\alpha>0$. This indicates that there is no rotational recovery after rapid photobleaching with circularly polarized light.


\end{document}